\documentclass[11pt,twoside]{article}

\usepackage{asp2006}
\usepackage{epsf}
\usepackage{psfig}
\usepackage{lscape}

\begin{document}
\title{Chemical and Spectrophotometric Evolutionary Models \\
for Emission Line Star-forming Galaxies}  
\author{Mart\'in-Manj\'on, M.L $^{1}$, Moll\'a, M.$^{2}$, 
D\'iaz, A.I $^{1}$ and Terlevich, R. $^{3}$} 
\affil{$^{1}$ UAM, Cantoblanco, Madrid 28049 , Spain \\
$^{2}$ CIEMAT, Avda.Complutense 22, Madrid 28040, Spain \\
$^{3}$ INAOE, Luis Enrique Erro 1, Tonantzintla, Puebla 72840, Mexico}

\begin{abstract} 

We present a self-consistent model under a star-bursting scenario for
H{\sc ii} galaxies, combining different codes of chemical evolution,
evolutionary population synthesis and photoionization. The results
obtained reproduce simultaneoulsy the observed abundances, diagnostic
diagrams and equivalent width-colour relations for local H{\sc ii} galaxies.

\end{abstract}

\section{Introduction}   

The current burst of star formation (SF) dominates the spectral energy
distribution (SED) in H{\sc ii} galaxies even if previous stellar
populations are present, making difficult to know the star formation
history (SFH) of the galaxy. In order to understand how the SF takes
place we study the viability of a model \citep{man08} using
simultaneously the whole available information for the galaxy sample:
the ionized gas, which defines the present time state of the galaxy,
and spectrophotometric parameters, related to the galaxy SFH.

We assume 11 successive attenuated star-bursts along 13.2 Ga in a
region with a total gas mass of 10$^{8}M_{\odot}$. In each burst a
certain amount of gas is consumed to form stars with a given initial
efficiency. The SFH and the age-metallicity relation are given by the
chemical evolution code \citep[based on][]{fer94}. The SED of the
ionizing population is computed using the single stellar populations
(SSP's) from \cite{mol00}. Finally, the emission lines are calculated
by the photoionization code CLOUDY \citep{fer98}.

\section{Main Model Results}

\begin{figure*}[!ht]
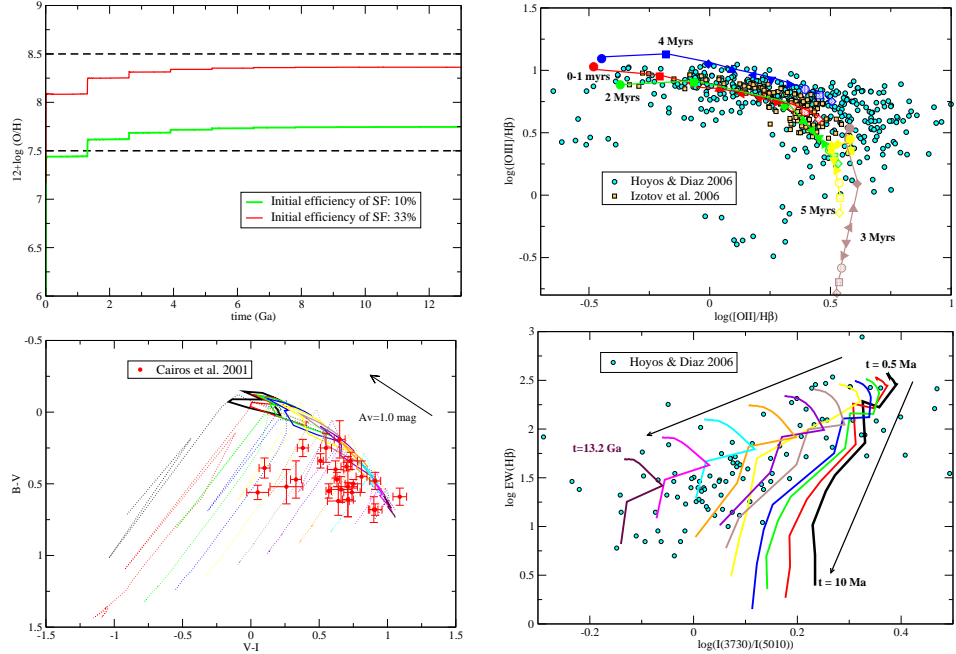

\epsfclipon 
\plottwo{fig1a.eps}{fig1b.eps}
\plottwo{fig1c.eps}{fig1d.eps}
\caption{a)Oxygen abundance with efficiencies $\epsilon= 0.3$ and 0.1. b)Diagnostic diagram,
c)Broad-band colors,d)Relation EW(H$\beta$) {\sl vs}
I(3729)/I(5010) ($\epsilon = 0.3$ model).}
\label{fig1}
\end{figure*}

The evolution of the oxygen abundance is shown in
Fig.~\ref{fig1}a. Models with initial efficiencies $\epsilon \sim $
10\% to 33\% reproduce the data range of H{\sc ii} galaxies,\citep[][dashed lines]{hoy06}.  
Once the SED of the ionizing
continuum is computed, we obtain the emission lines, giving
information about the current burst of SF.  Fig.~\ref{fig1}b shows an
excitation diagram for the ionized nebula. The model reproduces the
observational data for H{\sc ii} galaxies. The effect of an
underlying population should be more easily seen in the observed
colours of these galaxies, since any star-burst previous to the
currently observed one will contribute substantially to the total
continuum luminosity at the different wavebands. The broad band
colours with and without the contribution by the stronger emission lines are
shown in Fig.~\ref{fig1}c (dashed and solid lines respectively).


Most of observed EW(H$\beta$) values for H{\sc ii} galaxies are lower
than 150\AA\, which implies the existence of an old non ionizing
population. We plot the evolution of EW(H$\beta$) {\sl vs} the
pseudo-color I(3729)/I(5010) for the successive bursts of our model
with $\epsilon=0.3$ in Fig~\ref{fig1}d.  The data trend is reproduced
by our model, not by any SSP. A metal-poor SSP (black solid line)
show bluer colours than observed. In order to decrease EW(H$\beta$)
and obtain redder colours, a more metal-rich ($Z\sim Z_{\odot}$) SSP
might be selected, but such high abundance is not observed. According
to our model, a stellar population of at least 1.3Ga must contribute
to the continuum, making it redder and, simultaneously, decreasing
EW($\beta$).  In summary, our model reproduces abundances, colours,
and emission lines of H{\sc ii} galaxies.


\begin{thebibliography}{}

\bibitem[\protect\citeauthoryear{{Cair{\'o}s}, {Caon},
{V{\'{\i}}lchez}, et~al.}{{Cair{\'o}s} et~al.}{2001b}]{cai012}
{Cair{\'o}s} L.~M., {Caon} N., {V{\'{\i}}lchez} J.~M.,
et~al. 2001, ApJS, 136, 393

\bibitem[\protect\citeauthoryear{{Ferland}, {Korista}, {Verner},
et~al.}{{Ferland} et~al.}{1998}]{fer98}
{Ferland} G.~J., {Korista} K.~T., {Verner} D.~A., et.~al. 1998, PASP, 110, 761

\bibitem[\protect\citeauthoryear{{Ferrini}, {Moll\'{a}}, {Pardi} \&
{D{\'{\i}}az}}{{Ferrini} et~al.}{1994}]{fer94} {Ferrini} F.,
{Moll\'{a}} M., {Pardi} M.~C., {D{\'{\i}}az} A.~I., 1994, ApJ, 427,
745

\bibitem[\protect\citeauthoryear{{Hoyos} \& {D{\'{\i}}az}}{{Hoyos} \&
{D{\'{\i}}az}}{2006}]{hoy06} {Hoyos} C., {D{\'{\i}}az} A.~I., 2006,
MNRAS, 365, 454

\bibitem[\protect\citeauthoryear{{Izotov}, {Stasi{\'n}ska}, {Meynet},
{Guseva} \& {Thuan}}{{Izotov} et~al.}{2006}]{izo06} {Izotov} Y.~I.,
{Stasi{\'n}ska} G., {Meynet} G., et al. 2006,
A\&A, 448, 955

\bibitem[\protect\citeauthoryear{{Mart\'{\i}n-Manj\'{o}n}, {Moll\'{a}},
{D\'{\i}az} \& {Terlevich}}{{Mart\'{\i}n-Manj\'{o}n} et al.~}{2008}]{man08}
Mart\'{\i}n-Manj\'{o}n, M.L.,Moll\'{a}, M., D\'{\i}az, A.I. 
\& Terlevich, R. 2008, MNRAS, submitted

\bibitem[\protect\citeauthoryear{Moll{\'a} \& 
Garc{\'{\i}}a-Vargas}{2000}]{mol00} Moll{\'a} M.,
Garc{\'{\i}}a-Vargas M.~L., 2000, A\&A, 359, 18 

\end{thebibliography}
\end{document}